\documentstyle[prb,twocolumn,aps,epsfig]{revtex}
\begin{document}

\draft
\tighten

\title{Frustrated antiferromagnetic quantum spin chains
for spin length $S>1$}

\author{R. Roth} 
\address{Fachbereich Physik, Bergische Universit\"{a}t Wuppertal,
42097 Wuppertal, Germany}
\author{U. Schollw\"{o}ck}
\address{Sektion Physik, Ludwig-Maximilians-Univ.\ M\"{u}nchen,
Theresienstr.\ 37, 80333 Munich, Germany}

\date{April 30, 1998}

\maketitle
\begin{abstract}
We investigate frustrated antiferromagnetic Heisenberg quantum spin chains
at $T=0$ for
$S=\frac{3}{2}$ and $S=2$ using the DMRG method. We localize disorder and
Lifshitz points, confirming that quantum disorder points can be seen as
quantum remnants of classical phase transitions. 
Both in the $S=\frac{3}{2}$ and the $S=2$ chain, we observe 
the disappearance of effectively free $S=\frac{1}{2}$ and $S=1$ end spins
respectively. The frustrated spin chain is therefore a suitable system
for clearly showing the existence of free end spins $S'=[S/2]$ also in 
half-integer antiferromagnetic spin chains with $S>\frac{1}{2}$. 
We suggest that the first order transition 
found for $S=1$ in our previous work\cite{Kolezhuk 96} is present in {\em all}
frustrated spin chains with $S>\frac{1}{2}$, characterized by the disappearance
of effectively free end spins with $S'=[S/2]$. 
\end{abstract}

\narrowtext
\section{Introduction}
Because of its particularly simple structure the isotropic antiferromagnetic
(AFM) Heisenberg quantum spin chain with frustration through an AFM
next-nearest-neighbor (NNN) interaction at $T=0$,
\begin{equation}
{\cal H} = \sum_{i} {\bf S}_{i} \cdot {\bf S}_{i+1} + \alpha 
\sum_{i} {\bf S}_{i} \cdot {\bf S}_{i+2}, \mbox{\ \ \ } (\alpha>0),
\end{equation}
can be considered a textbook example for a system with competing interactions.
In recent years, this model has been studied extensively\cite{Diep 94}
because of its theoretical interest and experimental realizations. 
However, so far mainly results for spin lengths $S=\frac{1}{2}$ and $S=1$ are 
available\cite{Kolezhuk 96,Diep 94,Okamoto 92,Chitra 95,Bursill 95,White 96,Pati 96b}. Although these
cases are the most relevant ones for experiment, the underlying theory
cannot be considered complete; e.g., the generalization to large $S$ and the
emergence of the classical limit are unclear. In this paper, we will first 
outline the available results, and take a step towards large-$S$ behavior by 
numerically investigating the $S=\frac{3}{2}$ and $S=2$ case.
Our results, among them a clear demonstration of free end spins in the
$S=\frac{3}{2}$ chain, will allow us to extract the large-$S$
behavior and to propose a scenario
for frustrated chains with arbitrary $S$ accommodating available
results.

\section{$S=\frac{1}{2}$ and $S=1$ frustrated chains} 
Unfrustrated ($\alpha=0$) chains show strongly
different behavior for half-integer and integer $S$\cite{Haldane 83}; the
same holds if frustration is switched on.
 
The frustrated spin-$\frac{1}{2}$ chain is characterized by a critical
phase with AFM correlations for $\alpha<\alpha_{C}=0.2411$\cite{Okamoto 92}.
At $\alpha_{C}$, there is a continuous phase transition to a dimerized
gapped phase for all $\alpha>\alpha_{C}$. The dimerized phase 
is characterized by the
non-vanishing dimerization order parameter
\begin{equation}
D(\alpha)=| \langle {\bf S}_{i}\cdot{\bf S}_{i+1}-
{\bf S}_{i}\cdot{\bf S}_{i-1} \rangle | ,
\end{equation}
which reaches its maximum for $\alpha=0.5781$ and vanishes exponentially for
large $\alpha$\cite{White 96}.
The gap opens exponentially\cite{Haldane 82},
reaches a maximum at $\alpha\approx 0.6$, and disappears in the large-$\alpha$
limit, where the frustrated chain decomposes into two unfrustrated critical
Heisenberg chains\cite{White 96}. This scenario illustrates
the generalized Lieb-Schultz-Mattis theorem\cite{Affleck 86},
which allows just these two types of phases for a half-integer spin
chain with rotational and translational invariance. While the phase transition
does not affect the rotational invariance of the ground state, parity
is broken. For $S=\frac{3}{2}$, the transition
is suggested to be at $\alpha_C \approx 0.33$, and it can
be reasonably concluded that it generalizes to all half-integer 
spins\cite{Schulz 89}.

The frustrated spin-1 chain\cite{Kolezhuk 96} is gapped for all $\alpha$. For
$\alpha<\alpha_C=0.745$, one finds a phase which can be characterized by the
$S=1$ Affleck-Kennedy-Lieb-Tasaki (AKLT) model\cite{Affleck 87}. 
This phase has
non-vanishing nonlocal string order\cite{Nijs 89} and effectively free 
$S=\frac{1}{2}$ spins at the ends of open chains. These give rise to a 
low-lying excitation triplet, which is degenerate with the ground state in
the thermodynamic (TD) limit\cite{Kennedy 90}. While this excitation is clearly
a boundary effect, it is linked to bulk behavior, as shown by the existence
of nonlocal string order.
At $\alpha_C$, there is a first order transition characterized by discontinuous
disappearance of the nonlocal string order and the edge excitation
triplet\cite{Kolezhuk 96}. 
The large-$\alpha$ phase without string order or free end spins
can be understood by a next-nearest neighbor generalization of the AKLT-model
which can be extended to all $S$\cite{Kolezhuk 96}.

Both chains exhibit disorder points and Lifshitz points (i.e.\ points, where
the correlation function in real space becomes incommensurate, or where the
structure function develops a two-peak structure, respectively; for details
of the concept, see Refs.\ \onlinecite{Kolezhuk 96,Scholl 96a} and Refs.\ cited 
therein. Some authors merely call the tricritical point, where classical
disorder and Lifshitz lines meet, a Lifshitz point.). 
For $S=\frac{1}{2}$, the disorder point $\alpha_D=0.5$ is the 
well-known Majumdar-Ghosh point, the Lifshitz point is at 
$\alpha_L=0.5206$\cite{Bursill 95}. The respective results for $S=1$ are
$\alpha_D=0.284(1)$ and $\alpha_L=0.3725(25)$\cite{Kolezhuk 96}. We interpret
these as proposed in Refs.\ \onlinecite{Kolezhuk 96,Scholl 96a}:
In the classical frustrated chain, there is a $T=0$ phase transition from 
commensurate AFM order to incommensurate spiral order
for $\alpha_{cl}=0.25$. At
finite $T$, the classical chain is disordered, but should exhibit disorder
and Lifshitz lines, in line with the usual explanation of classical
disorder lines\cite{Stephenson 69}. The link between classical chains
at $T\neq 0$ and quantum chains at $T=0$ is established by the nonlinear
$\sigma$-model relation 
$T \propto 1/S$, which holds at least for small frustration. 
This scenario has three implications: First, disorder and
Lifshitz points are nothing but the quantum remnants of the classical phase
transition, second, they should thus be present for arbitrary $S<\infty$; and
third, in
the classical limit, their positions should converge towards each other and to
$\alpha_{cl}$\cite{Kolezhuk 96}. These implications remain to be tested.
Let it be remarked that as a side result, we conclude that the 
transitions in the $S=\frac{1}{2}$ and the $S=1$ chain
are pure quantum phenomena unrelated to the classical transition.

We apply now
the DMRG algorithm\cite{White 92} to frustrated $S=\frac{3}{2}$ and $S=2$ quantum spin chains,
considering chains of length up to 300 sites, keeping up to 300 states.
All chains considered have open boundary conditions. We map out the chain
behavior with $\alpha$. 
 
\section{Disorder points and Lifshitz points}
Numerically, we can identify disorder point and Lifshitz point in both cases
investigated, using the criteria already employed in Ref.\onlinecite{Kolezhuk 96}. 
For $S=\frac{3}{2}$, we find a disorder point $\alpha_D=0.386(1)$
and a Lifshitz point $\alpha_L=0.388(1)$; they are close, but separable. 
The correlation
length has a minimum $\xi(\alpha_D)\approx 1.9$. For $S=2$, disorder point and
Lifshitz point are found for $\alpha_L=0.289(1)$ and $\alpha_D=0.325(5)$, 
with a minimum of correlation length $\xi(\alpha_D)\approx 1.8$.

Haldane's now-well established conjecture\cite{Haldane 83} 
predicts a different behavior of half-integer and
integer unfrustrated Heisenberg spin chains. For reasons of continuity,
one can expect that this leads to substantial differences in not too strongly
frustrated half-integer and integer spin chains also. From this one can 
expect that the disorder and Lifshitz points of half-integer 
and integer spin chains take different tracks to the classical limit. 
This is supported by putting known results together (see Table 1).

Half-integer and integer spin chains show clearly different
behavior. For half-integer spin chains we find, that the disorder and
Lifshitz points, which were already relatively close together for 
$S=\frac{1}{2}$, are just distinguishable in the case of $S=\frac{3}{2}$ 
and much closer to $\alpha_{cl}$ than their $S=\frac{1}{2}$ counterparts. 
At the same time, the disorder point correlation length increases. 
As the $S=\frac{1}{2}$ disorder point is the Majumdar-Ghosh ground state,
only neighboring spins are correlated \cite{MG}, so that the correlation 
length is not defined or zero.

For the $S=2$ chain we find that disorder and Lifshitz points move, as
expected, closer to each other, but not as close as in the half-integer 
case, together with an increasing correlation length
at the disorder point. However, the disorder point of the $S=2$
chain is slightly further away from $\alpha_{cl}=\frac{1}{4}$ than for the
$S=1$ chain. This small exception is in our view due to the fact that our
model of the behavior of disorder and Lifshitz points does not take into
account other physical effects in the chains which may influence the exact
locations, but only the difference between half-integer and integer spin chains.
Once disorder and Lifshitz points have moved close to the classical limit,
these additional effects may be strong enough to enforce deviations from
the by then slow convergence.
Summarizing, we take these
results as strong support for our explanation of quantum disorder points
and the implications listed above.

\begin{figure}
\centering\epsfig{file=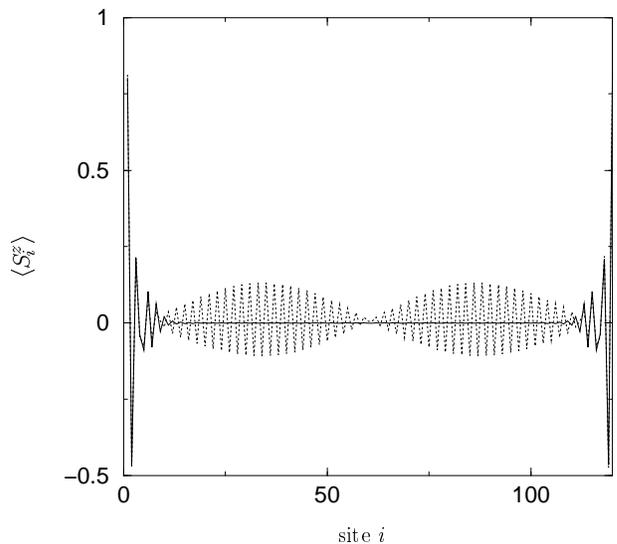,scale=0.6}
\vspace{0.3truecm}
\caption{\label{fig:kennedy32.385} Magnetization of the lowest-lying spin-1
(solid)
and spin-2 excitations (dashed) for the frustrated $S=\frac{3}{2}$ chain 
at $\alpha=0.385$ with $L=120$. 
The spin-1 excitation is clearly a boundary 
excitation (Kennedy triplet), the spin-2 excitation combines a bulk excitation
with the spin-1 edge excitation.}
\end{figure}

\begin{figure}
\centering\epsfig{file=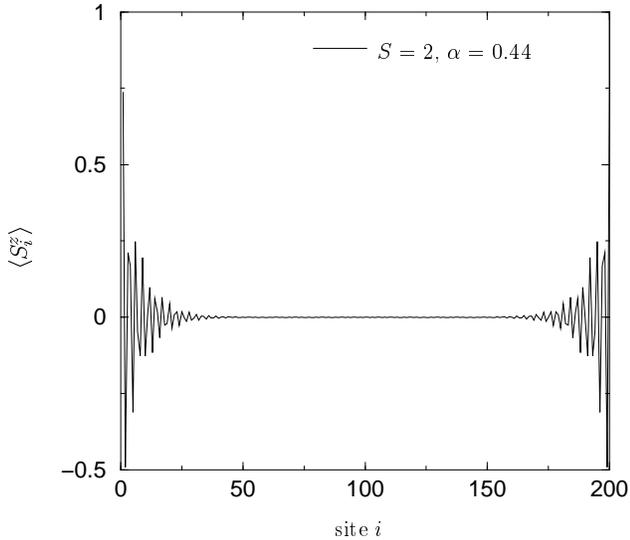,scale=0.6}
\vspace{0.3truecm}
\caption{\label{fig:kennedy2.44} Magnetization of the lowest-lying spin-1
excitation for the frustrated $S=2$ chain at $\alpha=0.44$ for $L=200$,
below the point of the disappearance of
edge excitations.}
\end{figure}

\section{Free end spins in open spin chains}
An important observation in the frustrated $S=1$ chain was the existence
of free $S=\frac{1}{2}$ end spins disappearing for sufficiently strong
frustration. No such observation can be made in the $S=\frac{1}{2}$ chain.
We observe, for weak frustration, the existence of free $s=\frac{1}{2}$
end spins in the $S=\frac{3}{2}$ chain, giving rise to a low-lying edge 
excitation triplet, well known as Kennedy triplet in $S=1$ chains. 
Ng\cite{Ng 94} has conjectured the existence of free end spins in all
unfrustrated open chains for $S\geq 1$. He derived a Berry phase contribution
to the effective action of the nonlinear $\sigma$-model, predicting free
end spins $S'=[S/2]$ for all $S\geq 1$.
In the
critical phase, edge and bulk excitations are hard to distinguish 
numerically\cite{Ng 95}; the presence of a gapped phase in the $S=\frac{3}{2}$
frustrated spin chain makes the identification very easy due to the finite
correlation length. As example,
we show in Figure \ref{fig:kennedy32.385} 
the magnetization of the $S^z_{total}=1$ state of the
spin-1 end excitation triplet for $\alpha=0.385$ as well as the magnetization
of the lowest spin-2 excitation, which is the boundary excitation plus a
bulk excitation. Similar observations can be made for $S=2$ (Fig.\
\ref{fig:kennedy2.44}), where one finds free $S=1$ spins, but similar
observations had already been clearly obtained for $S=2$ at the unfrustrated 
point\cite{Ng 95,Scholl 95,Nishiyama 95}.

\begin{figure}
\centering\epsfig{file=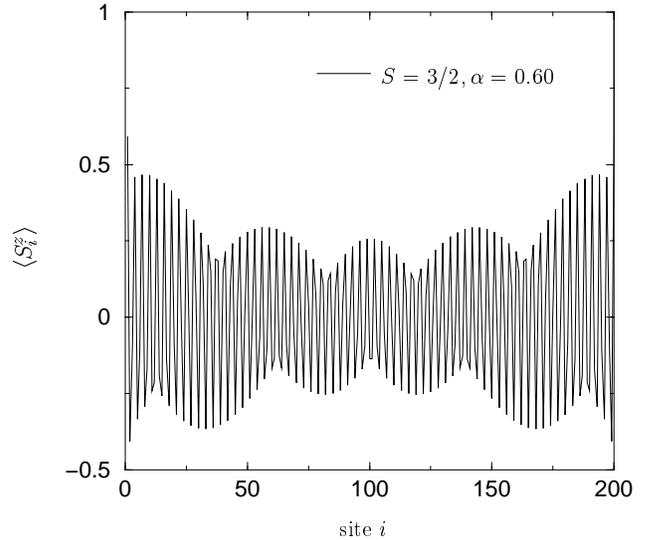,scale=0.6}
\vspace{0.3truecm}
\caption{\label{fig:kennedy32.60} Magnetization of the lowest-lying spin-1
excitation for the frustrated $S=\frac{3}{2}$ chain at $\alpha=0.60$,
which is a bulk excitation.}
\end{figure}

\begin{figure}
\centering\epsfig{file=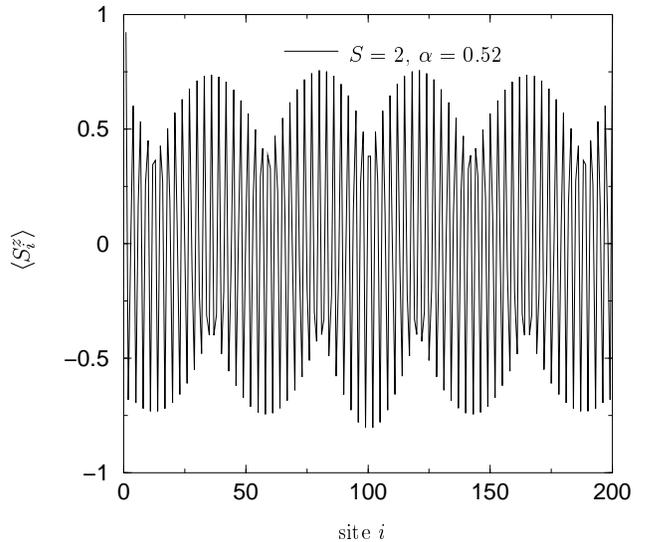,scale=0.6}
\vspace{0.3truecm}
\caption{\label{fig:kennedy2.52} Magnetization of the lowest-lying spin-1
excitation for the frustrated $S=2$ chain at
$\alpha=0.52$ for $L=200$ ($\approx 10\xi$), 
above the point of the disappearance of
edge excitations.}
\end{figure}

Furthermore, we observe that the free end spins disappear, as in the $S=1$
case, both for $S=\frac{3}{2}$ and $S=2$. The lowest excitations shown
in Figs.\ \ref{fig:kennedy32.60} and \ref{fig:kennedy2.52}
are clear bulk excitations.

As explanation for this result, 
we invoke the model proposed by Ng\cite{Ng 94}. His
derivation inherently assumes
local N\'{e}el order: in his Berry phase contribution, local N\'{e}el order
leads to alternating signs in a sum over local phase contributions, which
can be interpreted as an integral of the first derivative of these 
contributions,
such that merely the boundary terms survive. 
For continuity reasons, we may also expect this
assumption to hold for weak frustration. For sufficiently strong
frustration, this assumption fails, as we will find short-ranged
{\em spiral} correlation. The free end spins should then disappear
for sufficiently strong frustration.

\begin{figure}
\centering\epsfig{file=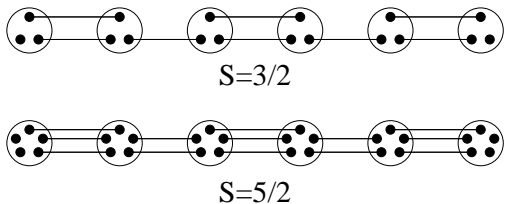,scale=0.6}
\vspace{0.3truecm}
\caption{\label{fig:toy} Valence bond toy model of the dimerized 
ground states of the $S=\frac{3}{2}$ and $S=\frac{5}{2}$ frustrated
chains. Open cirles are sites; full circles spin-$\frac{1}{2}$ fully
symmetrized on each site. Straight lines are singlet bonds. Note the
presence of free $S=\frac{1}{2}$ and $S=1$ spins at each end.}
\end{figure}

It might also be quite instructive to consider the
valence bond toy model of the dimerized 
ground states of the $S=\frac{3}{2}$ and $S=\frac{5}{2}$ frustrated
chains as shown in Figure \ref{fig:toy}. This toy model
cannot be constructed for $S=\frac{1}{2}$. It is basically a conventional
AKLT ground state, where the odd number of singlet bonds emerging from each
site is split such between right and left neighbor that the difference
between bond numbers is minimal, i.e.\ 1, arguing that the underlying 
interaction is non-dimerized. This gives alternating ``strong'' and
``weak'' bonds.
In an open chain of even length, there will be
``strong bonds'' at each end for minimal total energy, 
which leaves effectively free
spins $S'=[S/2]$. For this model, an argumentation as for integer spins can
be repeated to predict a first order transition for all half-integer 
$S\geq\frac{3}{2}$. Note that free end spins also exist if
resonating bonds restore criticality\cite{Affleck 87}. 

As in the $S\neq 1$ case the string order parameter is not characterizing the
disappearance of the free end spins, the precise localization of the point
of disappearance is difficult: in a small region, mixed bulk-boundary 
excitations will be energetically most favorable in finite length systems.
Considering the lowest excitation gap, we were able to narrow down the
points to $\alpha=0.48(2)$ for $S=\frac{3}{2}$ and $\alpha=0.46(1)$ for
$S=2$.

It is reasonable to assume that the observed scenario holds for all larger values of $S$ also.
\section{Critical and Massive Phases}
As for $S=\frac{1}{2}$, $S=\frac{3}{2}$ has a Kosterlitz-Thouless
phase transition from an antiferromagnetic critical phase to a dimerized one,
whereas the $S=2$ chain is massive for all values of frustration.

Therefore it is expected, that the spin gap in the $\frac{3}{2}$ chain shows close
to the transition the same behavior as the $\frac{1}{2}$ chain, 
namely\cite{Haldane 82}
\begin{equation} \label{Gapform}
\Delta(\alpha)\sim\frac{1}{\sqrt{\alpha-\alpha_c}}
\exp{(-\frac{const.}{\alpha-\alpha_c})}.
\end{equation}

We want to show with our DMRG results, that the gap of the $S=\frac{3}{2}$
chain opens according to (\ref{Gapform}) and use (\ref{Gapform}) to determine
$\alpha_{c,\frac{3}{2}}$. Before we present our results, let us first point out,
that because DMRG is a variational method and hence overestimates energies and
energy gaps, fitting DMRG data with (\ref{Gapform}) will return 
a $\alpha_{c,\frac{3}{2}}$ too small. Furthermore we
are restricted to a region $\alpha>\alpha_{c,\frac{3}{2}}$, for which the gap is
significant bigger than the numerical error. To test our procedure, 
we calculate the spin gap of the frustrated $S=\frac{1}{2}$ chain close to its
transition and used the known value of 
$\alpha_{c,\frac{1}{2}}=0.2411$\cite{Okamoto 92}. We find good agreement.

To determine the gap of the $S=\frac{3}{2}$ chain, we calculate the
energies of the ground state and the first excitation for chains of up to $L=300$
spins with up to $M=300$ block states kept in the range $\alpha=0.35\dots 0.385$.
We extrapolate $M\to\infty$ and $L\to\infty$ to get the gap in the thermodynamic
limit. $\alpha=0.35$ is the smallest value of $\alpha$ for which we can see a
significant gap. The gap changes its
behavior at approximately $\alpha=0.375$, so we fit our gap data in the range
$0.35\leq\alpha\leq 0.37$ with (\ref{Gapform}) and determine
$\alpha_{c,\frac{3}{2}}\approx 0.29$, a value that we consider as lower bound of the
true value. This is consistent with an earlier approximate 
result\cite{Schulz 89} 
$\alpha_{c,\frac{3}{2}}\approx 0.33$. 

\begin{figure}
\centering\epsfig{file=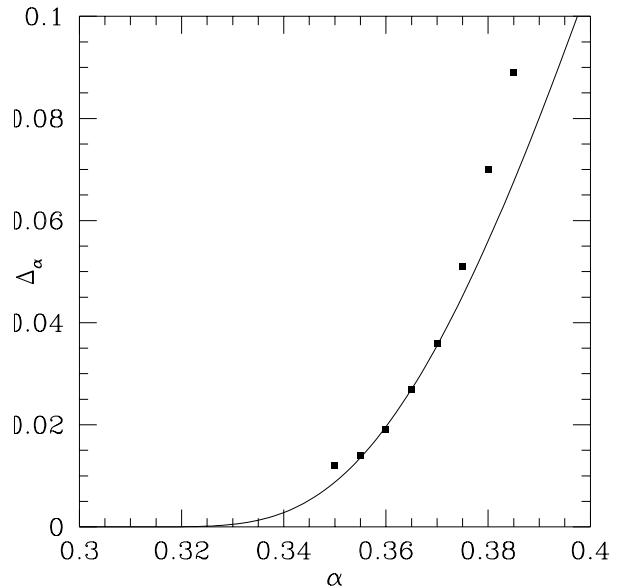,scale=0.4}
\vspace{0.3truecm}
\caption{\label{fig:gap32} Onset of the gap in the $S=\frac{3}{2}$
quantum spin chain at the Kosterlitz-Thouless transition.}
\end{figure}

\begin{figure}
\centering\epsfig{file=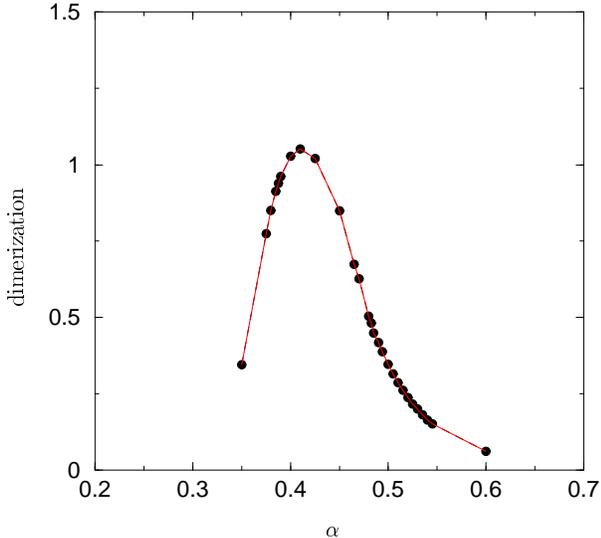,scale=0.6}
\vspace{0.3truecm}
\caption{\label{fig:dimer} Dimerization for the frustrated $S=\frac{3}{2}$
spin chain.}
\end{figure}

In Figure \ref{fig:gap32} we show the opening of the gap. Our DMRG results are
represented by dots, the solid line shows the fit. It clearly can be seen that
for $\alpha>0.37$ the gap changes its behavior. The smallest gap value we have is 
overestimated due to systematic DMRG errors. 
For large frustration, the gap decreases again monotonically.
The gapped phase is characterized by  finite dimerization (Figure
\ref{fig:dimer}), which reaches its maximum at $\alpha=0.415(5)$, for smaller
frustration than in the $S=\frac{1}{2}$ case, and then also drops
monotonically. While bigger in absolute terms than for $S=\frac{1}{2}$,
relative to the ground state energy it is weaker, as it should be: in the
classical limit, zero dimerization is expected, if the $S^{2}$ factor
governing the ground state energy is
scaled out. 

\begin{figure}
\centering\epsfig{file=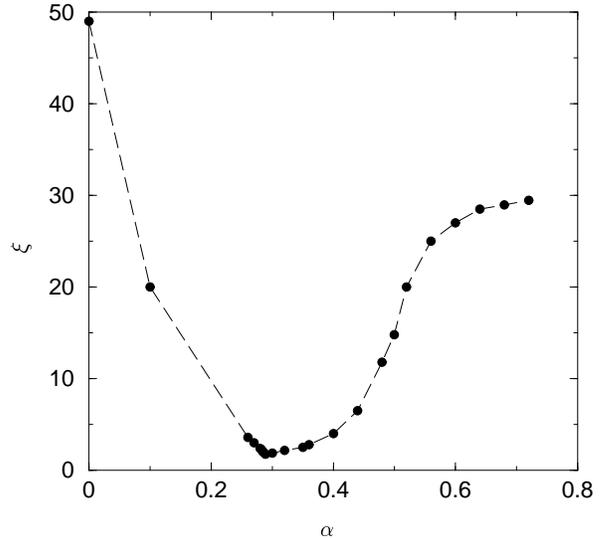,scale=0.6}
\vspace{0.3truecm}
\caption{\label{fig:xi} Correlation lengths for the frustrated $S=2$
spin chain. Values are systematically underestimated for large $\xi$ by
the DMRG. The saturation for large $\alpha$ is artificial, a slow convergence
to $\alpha\approx 100$ for $\alpha\rightarrow\infty$ being offset by
stronger underestimation due to a strong increase in the truncation error.}
\end{figure}

For $S=2$, we have determined the correlation length, which shows that the
chain is always massive. For large frustration, we expect that $\xi$ converges
to $2\xi(\alpha=0)$; the saturation effect in Figure \ref{fig:xi} is
purely artificial. It is well known that the DMRG underestimates correlation
lengths, the more so, if the DMRG
truncation error is large. For large frustration,
the chain starts to decompose into two chains, making the DMRG description
worse, dramatically increasing the truncation error. The saturation effect
comes from a growing underestimation of a growing correlation length.

The remaining important question is whether we can associate a
first order transition with the disappearance of the free end spins as 
in the $S=1$ case. We may expect that this transition, which is a pure
quantum effect, is softened for larger $S$. Moreover, we do not have a
clear indicator like the string order parameter in the $S=1$ case.
This, together with the increasing numerical difficulties for larger $S$,
made it impossible for us to definitely answer this question.  
As the spin correlation length is finite for all spin lengths considered at
the point of end spin disappearance, at least
a second-order transition can be excluded.

At the disappearance of the end spins, 
we observe a drop of the truncation error for a fixed
number of block states by more than an order of magnitude, as opposed
to much smoother change of the truncation error with $\alpha$ for all other
values of frustration. This might be an indicator that
a first-order phase transition has happened.
Furthermore, just below the transition, 
spin-spin correlations with large
spin-spin distances change drastically in character from
bulk behavior. As we calculate correlations for
spins symmetric around the chain center, the spins involved are close to the
chain ends, indicating a coexistence phenomenon or a metastability
problem of DMRG. 
After the disappearance of end spins, 
the effect disappears. All this is however highly indirect evidence only,
but would be consistent with a first-order transition. Also for systematic
reasons, we suggest that there is such a transition for all $S\geq 1$, but
further work
will be necessary to attack this question.

\section{Conclusion} 
In conclusion, a coherent picture of the behavior of AFM 
quantum spin chains with NNN frustration emerges. The wealth
of phenomena observed is due to a complex interplay of different features,
only some of which are sensitive to the difference between half-integer and
integer spins. Pure quantum phenomena coexist with traces of the classical
limit. The phase transition from AFM to spiral
order in the classical limit leaves quantum remnants in the form of disorder
and Lifshitz points, which converge towards the classical phase transition
point in the large-$S$ limit. This ``classical'' phenomenon is common to
all half-integer and integer spins. All transitions in the quantum
frustrated chains are therefore pure quantum phenomena and unrelated to the
classical limit.
Haldane's conjecture shows up in the
different fate of unfrustrated half-integer and integer spin chains, when
the frustrating interaction is switched on: whereas integer spin chains
remain gapped for all values of frustration, half-integer spin chains
exhibit a dimerising continuous phase transition from a critical 
antiferromagnetic phase to a dimerized gapped phase, a nice illustration of
the generalized Lieb-Schultz-Mattis theorem. The third prominent feature
is the disappearance of effectively free
$S'=[S/2]$ end spins 
which are present in all these chains with $S>\frac{1}{2}$ for low 
frustration. This
generalizes the previously found result in the
$S=1$ chain. Free end spins are thus also present in
half-integer open AFM spin chains. 
It was not possible to clearly show that there is also a first-order transition
present for $S>1$. While it would be systematic to expect this, it should
be kept in mind that the $S=1$ Heisenberg chain, among integer spin chains,
might be rather special\cite{Aschauer 98}. 

Most numerical
calculations were carried out on a 200 MHz PentiumPro running under Linux.

\begin{table}
\begin{tabular}{cccc} 
Spin length & $\alpha_D$ & $\alpha_L$ & $\xi_D$ \\ \tableline
$S=\frac{1}{2}$ & 0.5 & 0.52063(6) \cite{Burs} & --- \\ 
$S=1$ & 0.284(1) & 0.3725(25) & 1.20(2) \\ 
$S=\frac{3}{2}$ & 0.386(1) & 0.388(1) & 1.88(2) \\ 
$S=2$ & 0.289(1) & 0.325(5) & 1.78(2)\\ 
\end{tabular}

\caption{Location of disorder points $\alpha_D$ and Lifshitz points
$\alpha_L$, and the disorder point correlation length $\xi_D$ for
various spin lengths.}
\label{DOPTable}
\end{table}

\end{document}